\documentclass{article}

\usepackage[utf8]{inputenc} 
\usepackage[T1]{fontenc}    
\usepackage{url}            
\usepackage{booktabs}       
\usepackage{amsfonts}       
\usepackage{nicefrac}       
\usepackage{microtype}      
\usepackage{amsmath}
\usepackage{caption}
\usepackage{subcaption}
\usepackage{placeins}

\usepackage[numbers, compress]{natbib}
\usepackage{authblk}
\usepackage[a4paper, total={6.2in, 8.5in}]{geometry}

\usepackage{bm}             
\usepackage{amsmath}
\usepackage{amssymb}
\usepackage{graphicx}

\usepackage{titlesec}

\titlespacing\section{0pt}{8pt plus 4pt minus 2pt}{0pt plus 1pt minus 1pt}
\titlespacing\subsection{0pt}{8pt plus 4pt minus 2pt}{0pt plus 1pt minus 1pt}
\titlespacing\subsubsection{0pt}{8pt plus 4pt minus 2pt}{0pt plus 1pt minus 1pt}

\newcommand{\blank}{\mathrel{\;\cdot\;}}

\DeclareMathOperator*{\argmin}{argmin}

\begin{document}

\title{Uncovering solutions from  data corrupted by systematic errors: \\A physics-constrained convolutional neural network approach}

\author[1]{Daniel Kelshaw}
\author[1,2]{Luca Magri}

\affil[1]{Department of Aeronautics, Imperial College London}
\affil[2]{The Alan Turing Institute}

\date{}

\maketitle

\begin{abstract}
\noindent
Information on natural phenomena and engineering systems is typically contained in data. 
Data can be corrupted by systematic errors in models and experiments. 
In this paper, we propose a tool to uncover the spatiotemporal solution of the underlying physical system by removing the systematic errors from data. 
The tool is the physics-constrained convolutional neural network (PC-CNN), which combines information from both the system's governing equations and data.  
We focus on fundamental phenomena that are modelled by partial differential equations, such as linear convection, Burgers' equation, and two-dimensional turbulence. 
First, we formulate the problem, describe the physics-constrained convolutional neural network, and parameterise the systematic error. 
Second, we uncover the solutions from data corrupted by large multimodal systematic errors.  
Third, we perform a parametric study for different systematic errors.
We show that the method is robust.  
Fourth, we analyse the physical properties of the uncovered solutions. 
We show that the solutions inferred from the PC-CNN are physical, in contrast to the data corrupted by systematic errors that does not fulfil the governing equations. 
This work opens opportunities for removing epistemic errors from models, and systematic errors from measurements.
\end{abstract}

\section{Introduction}
%
Model estimates and experimental measurements can be affected by systematic errors. Systematic errors
can arise for a number of reasons: faulty experimental sensors~\cite{sciacchitano2015collaborative}, or low-fidelity numerical 
methods in which closing nonlinear multiscale equations can add bias, to name a few. The detection and removal of systematic errors 
has numerous applications, ranging from correcting biased
experimental observations, to enhancing results obtained from low-fidelity simulations~\cite{zucchelli2021multi}. In the field of fluid dynamics, 
experimental measurement of a flow-field is an inherently challenging process. Measurement techniques are often limited given the sensitivity of the 
flow to immersed sensors and probes, the introduction of which can fundamentally alter the intended behaviour of the system~\cite{adrian1991particle}. 
In many cases, non-intrusive methods such as particle-image-velocimetry are preferred, the results of which provide no guarantee of satisfying the underlying 
physical laws. Removing systematic errors that are present in these observations would yield the underlying solution of the system. Approaches for 
recovering a divergence-free field exist, but focus predominantly on filtering out small quantities of stochastic noise \cite{Vedula2005, Schiavazzi2014}.

Irrespective of the research domain, a contemporary issue in the field of modelling and simulation is the computational expense of running 
high-fidelity simulations~\cite{kochkov2021machine}. In many cases, practitioners rely on lower-fidelity methods, accepting assumptions or 
approximations, which invoke  degrees of model error. We posit that the obtained low-fidelity state can be considered as a corrupted observation because 
the underlying solution of the system subjected to a form of systematic error. Providing a mapping from the corrupted-state to the underlying solution would 
allow the inference of high-fidelity results given only low-fidelity data. In practice, we observe that verifying that  observations are 
characteristic of a given system is more straightforward than determining the solution itself. The former requires ensuring that the governing
equations are satisfied, whilst the latter requires a method to produce a candidate solution. The overarching goal of this paper is to design a method to remove large-amplitude systematic errors from data to uncover the true solution of the partial differential equation.

Systematic error removal in the literature typically considers the case of aleatoric noise-removal, which is a term attributed to methods that remove small, 
unbiassed stochastic variations from the underlying solution. This has been achieved through various methods, including filtering methods 
\cite{Kumar2017}; proper orthogonal decomposition \cite{Raiola2015, Mendez2017}; and the use of autoencoders \cite{Vincent2008}. 
Super-resolution can be considered a form of systematic error removal because its maps low-resolution observations to their equivalent high-resolution
representations. In this case, the super-resolution task seeks to introduce additional information, which are not present in the original observation. This 
resolves the solution at a higher spatial frequency, which is a subtly different task to removing errors from observations. Notable works in 
this area include methods based on a library of examples \cite{Freeman2002} and sparse representation in a library \cite{Yang2010}. 
More recent work has employed machine learning methods, including methods based on convolutional neural networks \cite{Dong2014}, and
generative adversarial methods \cite{Xie2018}.

By exploiting the universal function approximation property of neural networks, it is possible to obtain numerical surrogates for function
mappings or operators \cite[e.g.,][]{Hornik1989, Zhou2020}. For a network parameterised by weights, there exists an optimal 
set of weights that results in the desired mapping; the challenge being to realise these through an optimisation process. This 
optimisation is formulated as the minimisation of a loss function, the evaluation of which offers a distance metric to 
the desired solution. When considering forms of structured data, convolutional neural networks excel due to their ability to exploit 
spatial correlations in the data, which arise from the use of localised kernel operations. 
In the case of data-driven problems, a typical approach is to minimise a  measure of the distance between network realisations and 
the associated true value. In the absence of knowledge about the true state, it is possible to impose prior 
knowledge onto the system. For a physical problem, the classical means to accomplish this is through a form of 
physics-based regularisation, which introduces penalty terms in the loss function to promote physical solutions, i.e., penalising 
realisations that do not conform to the properties of the system \cite{Lagaris1998}. Inspired by~\citet{Lagaris1998}, \citet{Raissi2019} 
introduced physics-informed neural networks (PINNs), which replace  classical numerical solvers with surrogate models. The loss 
function leverages automatic differentiation to obtain numerical realisations of the gradients of the output (the 
predicted quantity) with respect to the input, which usually contains the spatiotemporal coordinates. 

In this work, we propose a method for removing systematic errors with a physics-constrained machine learning approach. Imposing prior
knowledge of the physics allows us to realise a mapping from the corrupted-state to the true-state in the form
of a convolutional neural network, as shown in Figure \ref{fig:overview}. We emphasise that this is not a PINN approach~\cite{Raissi2019} because we do not leverage automatic differentiation to obtain gradients of outputs with respect to inputs to constrain the physics. This 
network employs a time-batching scheme to effectively compute the time-dependant residuals without resorting to recurrent-based network 
architectures. Realisations of the network that do not conform to the given partial differential equation are penalised, which 
ensures that the network learns to produce physical predictions. Ultimately, this allows the effective removal of additive systematic error
from data.

In section §\ref{sec:defining_corruption_removal}, we formulate the problem of the systematic error removal. 
An overview of convolutional neural networks is provided in section §\ref{sec:cnn}, which highlights how we can exploit the spatial invariance and
describes the architecture employed for the systematic error removal task. We detail the methodology in section §\ref{sec:methodology}, which provides  
justifications for the design of the loss function, before  showcasing results in section §\ref{sec:results}. We provide a form of 
parameterised systematic error using a multimodal function~\cite{Rastrigin1974}, which allows us to explore the effect of high wavenumbers 
and magnitudes of the systematic error. Results are obtained from three systems of increasing physical complexity: linear convection-diffusion,
nonlinear convection-diffusion, and a two-dimensional turbulent flow. We further analyse results for the two-dimensional turbulent flow case,
investigating the physical coherence of network predictions. Section~§\ref{sec:conclusions} ends the paper. 

\section{Problem formulation} \label{sec:defining_corruption_removal}
%
We consider physical systems that can be modelled by partial differential equations (PDEs). Upon suitable spatial discretisation, a PDE
with boundary conditions is viewed as a dynamical system in the form of 
\begin{equation} \label{eqn:dynamical_system}
   \mathcal{R}({\bm u}; \lambda) \equiv \partial_t {\bm u} - \mathcal{N}({\bm u}; \lambda),
\end{equation}
where ${\bm x} \in \Omega \subset \mathbb{R}^n$ denotes the spatial location; $t \in [0, T] \subset \mathbb{R}_{\geq 0}$ is the time; 
${\bm u}: \Omega \times [0, T] \rightarrow \mathbb{R}^m$ is the state; $\lambda$ are physical parameters of the system; $\mathcal{N}$ is a 
sufficiently smooth differential operator; $\mathcal{R}$ is the residual; and $\partial_t$ is the partial derivative with respect to time.
A solution of the PDE ${\bm u}^\ast$ is the function that makes the residual vanish, i.e., $\mathcal{R}({\bm u}^\ast; \lambda) = 0$.

\begin{figure}[ht]
    \centering
    \includegraphics[width=0.75\linewidth]{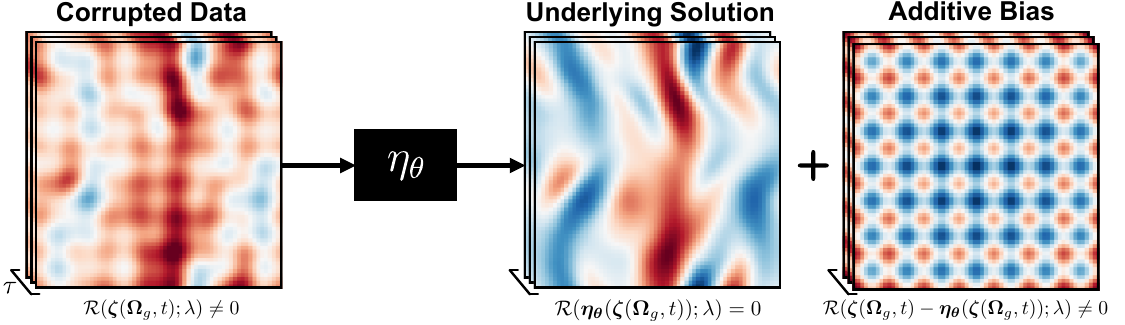}
    \caption{
        Removal of systematic error from data. The model ${\bm {\eta_\theta}}$ is responsible for recovering the underlying
        solution, ${\bm u}^\ast({\bm \Omega}_g, t)$, from the corrupted data, ${\bm \zeta}(\bm{\Omega}_g, t)$. The systematic error, ${\bm \phi}({\bm x})$, is the difference between the corrupted data and the underlying solution.
    }
    \label{fig:overview}
\end{figure}

We consider that the observations on the system's state, ${\bm \zeta}({\bm x}, t)$, are corrupted by an additive systematic error
\begin{equation}
    \bm{\zeta}(\bm{x}, t) = {\bm u}^\ast(\bm{x}, t) + {\bm \phi}({\bm x}), \label{eq:corruption_definition}
\end{equation}
where ${\bm \phi}$ is the stationary systematic error, which is spatially varying. For example, additive systematic errors can be caused by miscalibrated sensors~\cite{sciacchitano2015collaborative}, 
or modelling assumptions~\cite{novoa2022real}. The quantity in Eq.~\eqref{eq:corruption_definition} constitutes the corrupted dataset,
which fulfils the inequality 
\begin{align}
    \mathcal{R}({\bm \zeta}({\bm x}, t); \lambda) \neq 0. 
\end{align}
Given the corrupted state, ${\bm \zeta}$, we wish to uncover the underlying solution to the governing equations, ${\bm u}^\ast$, which is
referred to as the true state. Mathematically, we need to compute a mapping, ${\bm {\eta_\theta}}$ such that 
\begin{equation}
    {\bm {\eta_\theta}}: {\bm \zeta}({\bm \Omega}_{g}, t) \mapsto {\bm u}^\ast({\bm \Omega}_{g}, t),
\end{equation}
where the domain $\Omega$ is discretised on a grid ${\bm \Omega}_{g} \subset \mathbb{R}^{N^n}$, which is uniform and structured 
in this study. We assume the mapping ${\bm {\eta_\theta}}$ to be a parametric function, which depends on the parameters ${\bm \theta}$ 
that need to be found. In section §\ref{sec:cnn}, we describe the parametric function  that we employ in this study. Figure \ref{fig:overview}
provides an overview of the systematic error removal problem, which is applied to a turbulent flow. The corrupted field ${\bm \zeta}(\bm{\Omega_g}, t)$ 
is passed as input to the appropriately trained model ${\bm {\eta_\theta}}$, which allows us to uncover the underlying solution ${\bm u^\ast}$ to the 
partial differential equation, and, as a byproduct, the additive systematic error ${\bm \phi}$.

The linear vs. nonlinear nature of the partial differential equations has a computational implication. 
For a linear partial differential equation, the residual is an explicit function of the systematic error, i.e., 
$\mathcal{R}({{\bm \zeta}({\bm x}, t)}; \lambda) = \mathcal{R}({{\bm \phi}({\bm x}, t)}; \lambda)$. This does not hold true
for nonlinear systems, which makes the problem of removing systematic error more challenging. This is discussed in section §\ref{sec:results}.

\FloatBarrier
\section{The Physics-constrained convolutional neural network} \label{sec:cnn}
We propose the physics-constrained convolutional neural network (PC-CNN) for uncovering solutions of PDEs from corrupted data. 
Convolutional neural networks are suitable tools for structured data, which ensure shift invariance~\cite{lecun2015deep, Gu2018}. 
When working with data from partial differential equations, the spatial structure is physically correlated with respect the characteristic
spatial scale of the problem, for example, correlation lengths in turbulent flows, diffusive spatio-temporal scales in convection, among 
others~\cite{pope2000turbulent}. By leveraging an architectural paradigm that exploits these spatial structures, we propose parameterised 
models that, when appropriately trained and tuned, can generalise to unseen data~\cite{racca2022modelling, doan2021}. We provide a brief 
summary of convolutional neural networks and refer the reader to \cite{magri2023notes} for a pedagogical explanation. A convolutional layer
is responsible for the discrete mapping 
\begin{equation}
    \kappa: ({\bm w}, {\bm b}, {\bm x}) \mapsto {\bm x} \ast {\bm w} + {\bm b},
\end{equation}
where ${\bm w} \in \mathbb{R}^{k^d \times c_o}$ represents a trainable kernel; ${\bm b} \in \mathbb{R}^{c_o}$ is an additive bias; 
$\ast$ the convolution operation; and  $\bm{x} \in \mathbb{R}^{m^d \times c_i}$ is the input data.  The number of parameters in the network 
is proportional to the dimensionality  of the kernels, whose spatial extent is determined by $k$, while the number of filters,
or channels, is given by $c_o$.  The dimensionality of the input is independent of the kernel and bias, with $c_i$ determining the number of 
input channels. The trainable kernel $\kappa$ operates locally around each pixel, which leverages information from the surrounding grid cells. 
As such, convolutional layers are an excellent choice for learning and exploiting local spatial correlations~\cite{lecun2015deep, Murata2020, Gao2021}. 
Each channel in the kernel is responsible for extracting different features present in the input data. This 
determines the degree to which an arbitrary function can be learned, as per the universal approximation theorem \cite{Hornik1989}. 
A convolutional neural network is an architecture that leverages multiple  convolutions, which can be viewed
as a composition of multiple layers where $Q$ is the number of layers
\begin{equation}
    {\bm {\eta_\theta}} = {\bm \eta}^{Q}_{\bm \theta_Q} \circ \cdots \circ h({\bm \eta}^{1}_{\bm \theta_1}) \circ h({\bm \eta}^{0}_{\bm \theta_0}).
\end{equation}
In the most basic form, each layer ${\bm \eta}_{\theta_i}^{i}$ is a convolution $\kappa$ followed by an element-wise nonlinear activation function
$h$. In the absence of nonlinearities, a network is capable only of learning a linear transformation, which restricts the function approximation 
space~\cite{Bishop1994}. The introduction of these nonlinear activations is a key ingredient to the expressivity of the network. The final layer ${\bm \eta}_{\bm \theta_Q}^{Q}$ 
is exempt from activation as this limits the expressive power of the output. There are a number of modifications that can be made to each layer, 
we refer the reader to~\cite{Aloysius2017} for a more general treatment of the topic.
As the filter is convolved over each discrete pixel, the operation on large spatial domains may become computationally expensive due to 
the $\mathcal{O}(k^2)$ scaling. To mitigate the increased computational expense, we choose a kernel size $k = 3$ for each convolutional layer, 
which is a common choice for CNN architectures~\cite{Aloysius2017}.

\begin{figure}[ht]
    \centering
    \includegraphics[width=\linewidth]{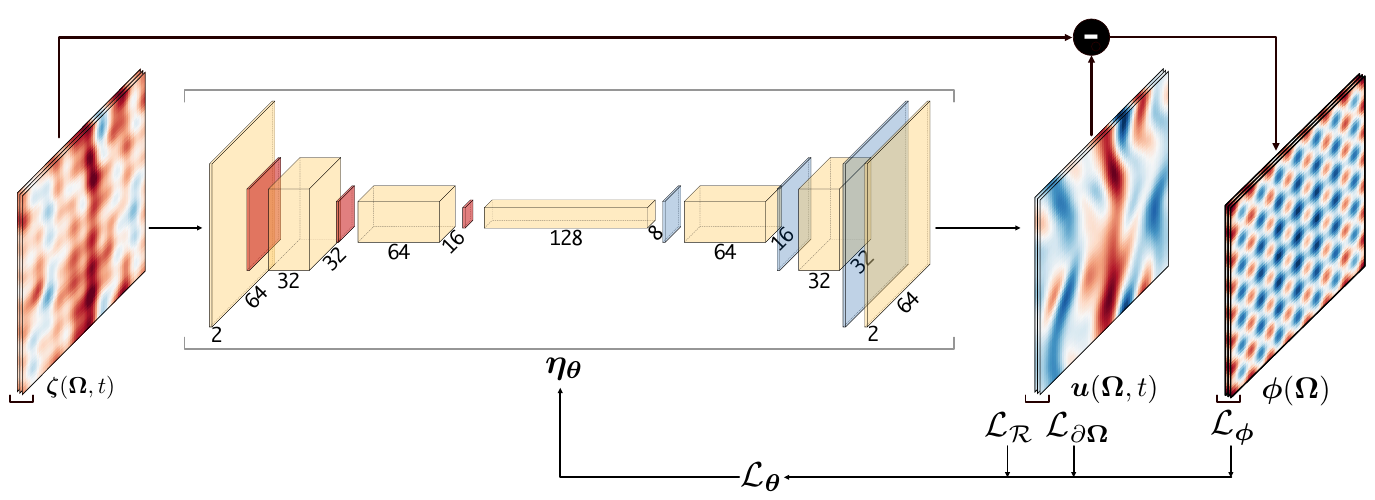}
    \caption{
        Diagram of network architecture and training pipeline. In the network architecture ${\bm {\eta_\theta}}$: yellow layers denote convolution operations; 
        red layers denote mean-pooling; blue layers denote bi-linear interpolation. Composite losses $\mathcal{L}_{\mathcal{R}}, \mathcal{L}_{\partial {\bm \Omega}}, \mathcal{L}_{\bm \phi}$
        are computed on the output field and inferred systematic error. These are then combined to compute the training loss $\mathcal{L}_{\bm \theta}$, which is used
        to update weights ${\bm \theta}$ of the network. We provide an in-depth explanation of these losses in section §\ref{sec:methodology}.
    }
    \label{fig:network_architecture}
\end{figure}

The basic architecture is extended through the use of pooling and upsampling layers to vary the spatial dimensionality of the input to sequential
convolutional layers. In this work, we employ mean pooling and bilinear upsampling with a kernel size of $k = 2$ in each case. Varying the dimensionality
of the input to the layers aids with feature extraction and provides another means to reduce the computational expense of the network -- each
filter $\kappa$ is convolved over a smaller spatial domain.
Figure \ref{fig:network_architecture} depicts the architecture used for the systematic error removal task, which highlights the operations 
employed with the dimensionality at each layer. The corrupted observations ${\bm \zeta}({\bm \Omega}_g, t)$ are passed as input to the 
network, with the outputs estimating the predicted true-state, ${\bm {\eta_\theta}}({\bm \zeta}({\bm \Omega}_g, t))$. 
\FloatBarrier
\subsection{Constraining the physics} \label{sec:methodology}
The convolutional neural network, ${\bm {\eta_\theta}}$, defines a nonlinear parametric mapping, whose parameters, ${\bm \theta}$,
are found by training. The information constrained in the training originates from two sources. First the data, which may be corrupted; 
and second, the prior knowledge on the physical system, which is encoded in partial differential equations. The training is an minimisation 
problem of a cost functional $\mathcal{L}_{\bm \theta}$, which reads  
\begin{equation} \label{eqn:optimization_problem}
\begin{aligned}
    {\bm \theta}^{*} &= \argmin_{\bm \theta} \mathcal{L}_{\bm \theta}, \\
    \mathcal{L}_{\bm \theta} &= \mathcal{L}_{\mathcal{R}} + \alpha \left( \mathcal{L}_{\partial {\bm \Omega}} + \mathcal{L}_{\bm \phi} \right),
\end{aligned}
\end{equation}
where $\alpha$ is a fixed, empirical weighting factor, which determines the relative importance of the loss terms.
Given corrupted-data ${\bm \zeta}({\bm \Omega_g}, t)$ at times $t \in \mathcal{T} \subseteq [0, T]$,
we define each of the loss terms in Eq.~\ref{eqn:optimization_problem} as 
\begin{align}
    \mathcal{L}_{\mathcal{R}} &= \frac{1}{N_{\mathcal{T}}} \sum_{t \in \mathcal{T}} \big\lVert
        \mathcal{R}\big({\bm {\eta_\theta}}({\bm \zeta}({\bm \Omega}_{g}, t)); \lambda \big)
   \big\rVert_{{\bm \Omega}_{g}}^{2}, \\
   \mathcal{L}_{\partial {\bm \Omega}} &= \frac{1}{N_{\mathcal{T}}} \sum_{t \in \mathcal{T}} \big\lVert
        {\bm {\eta_\theta}}({\bm \zeta}(\partial {\bm \Omega}_g, t)) - {\bm u}(\partial {\bm \Omega}_g, t)
    \big\rVert_{\partial {\bm \Omega}_g}^{2}, \\
   \mathcal{L}_{\bm \phi} &= \frac{1}{N_{\mathcal{T}}} \sum_{t \in \mathcal{T}} \big\lVert
        \partial_t \big[ {\bm \zeta}({\bm \Omega}_{g}, t) - {\bm {\eta_\theta}}({\bm \zeta}({\bm \Omega}_{g}, t)) \big]
   \big\rVert_{{\bm \Omega}_{g}}^{2},
\end{align}
where $N_{\mathcal{T}}$ is the number of time steps; $\partial {\bm \Omega}_g$ denotes boundary points of the grid, and 
$\lVert \blank \rVert_{{\bm \Omega}_g}$ is the $\ell^2$-norm over the given domain. The terms $\mathcal{L}_{\mathcal{R}}, \mathcal{L}_{\partial {\bm \Omega}}, 
\mathcal{L}_{\bm \phi}$ denote the residual loss, boundary loss, and systematic error loss, respectively.

In order to find an optimal set of parameters ${\bm \theta}^{*}$, the loss is designed to penalise, or regularise,
predictions, which do not conform to the desired output. We impose prior-knowledge of the dynamical system using the residual-based loss
$\mathcal{L}_{\mathcal{R}}$, promoting parameters that yield predictions that conform to the governing equations. Successful minimisation
of the residual-based loss ensures that network realisations satisfy the residual of the system, as shown in Eq.~\eqref{eqn:dynamical_system}.
The magnitude of parameter updates $\nicefrac{\partial \mathcal{L}_{\bm \theta}}{\partial {\bm \theta}}$ is proportional to the magnitude
of the loss, which ensures large violations of the residual are heavily regularised.

The residual-based loss alone is insufficient for obtaining the desired mapping. In the absence of collocation points the mapping ${\bm {\eta_\theta}}$
provides a non-unique mapping: any prediction that satisfies the governing equations is valid. To avoid this we impose knowledge of the
ground truth solution at select collocation points, chosen to lie on the boundaries of the domain $\partial {\bm \Omega}$. The data-driven boundary 
loss $\mathcal{L}_{\partial {\bm \Omega}}$ is designed to minimise the error between predictions and measurements at the chosen
collocation points. This has the effect of restricting the function approximation space to provide a unique mapping and ensuring that the
predictions satisfy the observations.

Inclusion of the systematic-error-based loss $\mathcal{L}_{\bm \phi}$ embeds our assumption of stationary systematic error ${\bm \phi} = {\bm \phi}(x)$.
Penalising network realisations that do not provide a unique systematic error helps stabilise training and drive predictions away from
trivial solutions, such as the stationary solution ${\bm u}({\bm \Omega}_{g}, t) = 0$. Consequently, this eliminates challenging local minima
in the loss surface, which improves convergence of the optimisation approach shown in Eq.~\eqref{eqn:optimization_problem}.

\subsection{Time-batching} \label{sec:time-batching}
The computation of the residual of partial differential equations in time is inherently temporal in nature, as can be seen in 
Eq.~\ref{eqn:dynamical_system}. Convolutional neural networks do not consider the sequentiality of the data in their architecture. In this 
paper, we propose time-batching the data to compute the time derivative $\partial_t {\bm u}$, which is necessary in the residual computation. 
We consider each sample passed to the network to constitute a window of $\tau$ sequential timesteps. This allows both computing predictions 
in parallel and sharing weights, which decreases the computational cost of the problem. Employing a time windowing approach 
provides a number of benefits, i.e., it allows us to process non-contiguous elements of the timeseries, and sample different points in 
the temporal domain. We provide an explicit formulation for the residual-based loss
\begin{equation}
    \mathcal{L}_{\mathcal{R}} = \frac{1}{\tau N_{\mathcal{T}}} \sum_{t \in \mathcal{T}} \sum_{n=0}^{\tau}
    \big\lVert 
        \partial_{t} {\bm {\eta_\theta}}({\bm \Omega}_{g}, t + n \Delta t))
    - 
        \mathcal{N}({\bm {\eta_\theta}}({\bm \zeta}({\bm \Omega}_{g}, t + n \Delta t)); \lambda)
    \big\rVert_{{\bm \Omega}_{g}}^{2},
\end{equation}
in which the time-derivative is computed using the explicit forward-Euler method over predictions at subsequent timesteps, 
and the differential operator is evaluated at each timestep. In so doing, we are able to obtain the residual for the predictions in a 
temporally local sense; rather than computing the residual across the entire time-domain, we are able to compute the residual over each
time window. This reduces the computational cost and increases the number of independent batches provided to the network.

We augment the other loss terms to operate in the same manner, operating over the time window as well as the conventional batch dimension.
While the time-independent boundary loss $\mathcal{L}_{\partial {\bm \Omega}}$ need not leverage the connection between adjacent timesteps, 
the systematic error loss $\mathcal{L}_{\bm \phi}$ computes the time-derivative in the same manner - evaluating the forward-Euler method between 
consecutive timesteps. 

\subsubsection{Choice of the time window}
The choice of the time window $\tau$ is important. For a fixed quantity of training data, we need to consider the trade-off between 
computing the residual over longer time windows, and sampling a larger number of windows from within the time domain. If we consider 
$N = \tau N_{\mathcal{T}}$ training samples, a larger value of $\tau$ corresponds to fewer time windows. Limiting the number of 
time windows used for training  has an adverse effect on the ability of the model to generalise; the information content of 
contiguous timesteps is scarcer than taking timesteps uniformly across the time domain. A further consideration is that, 
although evaluating the residual across large time windows promotes numerical stability, this  smooths the gradients 
computed in backpropagation, and makes the model more difficult to train, especially in the case of chaotic systems. 
Empirically, we choose $\tau = 2$, the minimum window size feasible for computing the residual-based loss $\mathcal{L}_{\mathcal{R}}$. 
We find that this is sufficient for training the network whilst simultaneously maximising the number of independent samples used for training. 
To avoid duplication of data in the training set, we ensure that all samples are at least $\tau$ timesteps apart so that independent time windows
are guaranteed to not contain any overlaps. 

\section{Data generation} \label{sec:solver}
A high-fidelity numerical solver provides the datasets used throughout this paper. We utilise a differentiable pseudospectral spatial 
discretisation to solve the partial differential equations of this paper~\cite{kolsol2022}. The solution is computed on on the spectral
grid ${\bm {\hat \Omega}}_k \in \mathbb{Z}^{K^n}$. This spectral discretisation spans the spatial domain 
${\Omega} \in [0, 2\pi) \subset \mathbb{R}^{2}$, enforcing periodic boundary conditions on $\partial \Omega$. A solution is produced by 
time-integration of the dynamical system with the explicit forward-Euler scheme, in which we choose the timestep $\Delta t$ to satisfy 
the Courant-Friedrichs-Lewy (CFL) condition~\cite{Canuto1988}. The initial conditions in each case are generated using the equation
\begin{equation} \label{eqn:initial_condition}
    {\bm u}^\ast({\bm {\hat \Omega}}_k, 0) = \frac{\iota e^{2\pi i {\bm \epsilon}}}{\sigma \sqrt{2\pi}} e^{-\frac{1}{2} \left( \frac{\lvert {\bm {\hat \Omega}}_k \rvert}{\sigma} \right)^2}
        \qquad \text{with} \quad {\bm \epsilon}_i \sim N(0, 1),
\end{equation}
where $\iota$ denotes the magnitude; $\sigma$ denotes the standard deviation; and ${\bm \epsilon}_i$ is a sample from a unit normal
distribution. Equation~\ref{eqn:initial_condition} produces a pseudo-random field scaled by the wavenumber ${\bm {\hat \Omega}}_k$,
which ensures that the resultant field has spatial structures of varying lengthscale. We take $\iota = 10, \sigma = 1.2$ for all simulations in
order to provide an initial condition, which provides numerical stability.

As a consequence of the Nyquist-Shannon sampling criterion~\cite{Canuto1988}, the resolution of the spectral grid ${\bm {\hat \Omega}}_k \in \mathbb{Z}^{K^n}$ 
places a lower bound on the spatial resolution. For a signal containing a maximal frequency $\omega_\text{max}$, the sampling 
frequency $\omega_s$ must satisfy the condition $\omega_\text{max} < \nicefrac{\omega_s}{2}$, therefore, we ensure that the spectral resolution 
satisfies $K < \nicefrac{N}{2}$. Violation of this condition induces spectral aliasing, in which energy content from frequencies exceeding the 
Nyquist limit $\nicefrac{\omega_s}{2}$ is fed back to the low frequencies, which amplifies energy content unphysically. 
To prevent aliasing, we employ a spectral grid ${\bm {\hat \Omega}}_k \in \mathbb{Z}^{32 \times 32}$, sampling on the physical grid 
${\bm \Omega_g} \in \mathbb{R}^{64 \times 64}$. Approaching the Nyquist limit allows us to resolve the smallest turbulent structures 
possible without introducing spectral aliasing.

The pseudospectral discretisation~\cite{kolsol2022} provides an efficient means to compute the differential operator $\mathcal{N}$. 
This allows us to accurately evaluate the residual-based loss $\mathcal{L}_{\mathcal{R}}$ in the Fourier domain
\begin{equation}
    \mathcal{L}_{\mathcal{R}} = \frac{1}{\tau N_{\mathcal{T}}} \sum_{t \in \mathcal{T}} \sum_{n=0}^{\tau} \lVert
        \partial_t \hat{{\bm {\eta_\theta}}}({\bm \zeta}({\bm \Omega}_{g}, t + n \Delta t)) - \hat{\mathcal{N}}(\hat{{\bm {\eta_\theta}}}({\bm \zeta}({\bm \Omega}_{g}, t + n \Delta t)))
    \rVert_{\hat{\bm \Omega}_{k}}^{2},
\end{equation}
in which $\hat{{\bm {\eta_\theta}}} = \mathcal{F} \circ {\bm {\eta_\theta}}$ where $\mathcal{F}$ is the Fourier  operator, and $\hat{\mathcal{N}}$ 
denotes the Fourier-transformed  differential operator. Computing the loss $\mathcal{L}_{\mathcal{R}}$ in the Fourier domain provides two advantages: 
(i) periodic boundary conditions are enforced automatically, which enables us to embed prior knowledge in the loss calculations; and 
(ii) gradient calculations have spectral accuracy~\cite{Canuto1988}.
In contrast, a conventional finite differences approach requires a computational stencil, the spatial extent of which places an error bound 
on the gradient computation. This error bound itself a function of the spatial resolution of the field.

\section{Results} \label{sec:results}
In this section, we first introduce the mathematical parameterisation of the systematic error and provide an overview of the numerical
details, data and performance metrics used in the study.
Second, we demonstrate the ability of the model to recover the underlying solution for three partial differential equations, and 
analyse the robustness and accuracy. We consider three partial differential equations of increasing complexity: the linear convection-diffusion, 
nonlinear convection-diffusion (Burgers' equation~\cite{bateman1915, burgers1948}), and two-dimensional turbulent Navier-Stokes equations
(the Kolmogorov flow~\cite{Fylladitakis2018}). 
Finally, we analyse the physical consistency of physics-constrained convolutional neural network (PC-CNN) predictions for the two-dimensional
turbulent flow. These analyses show the method, up to challenging partial differential equations exhibiting nonlinear, 
chaotic behaviour. 

\subsection{Parameterisation of the systematic error}
The systematic error is defined with the modified Rastrigin parameterisation~\cite{Rastrigin1974}, which is commonly used in non-convex
optimization benchmarks. The Rastrigin parameterisation by both frequency and magnitude allows us to assess the performance for a range of 
corrupted fields, which have multi-modal features. Multi-modality is a significant challenge for uncovering the solutions of partial 
differential equations from  corrupted data. We parameterise the systematic error as 
\begin{equation} \label{eqn:corruption}
    {\bm \phi}({\bm x}; \mathcal{M}, k_\phi, u_{\text{max}}) = 
        \frac{\mathcal{M} u_{\text{max}}}{2\pi^2 + 40} \left( 20 + \sum_{i=1}^{2} \left[ ({\bm x}_{i} - \pi)^{2} - 10 \cos(k_\phi ({\bm x}_{i} - \pi)) \right] \right)
\end{equation}
where $\mathcal{M}$ is the relative magnitude of systematic error; $u_{\text{max}}$ is the maximum velocity observed in the flow; and $k_\phi$ is the 
Rastrigin wavenumber of the systematic error. Parameterising the systematic error in this manner allows us to evaluate the performance 
of the methodology with respect to the magnitude and modality of the error.

\subsection{Numerical details, data, and performance metrics} \label{sec:overview_results}
We uncover the solutions of partial differential equations from corrupted data for three physical systems of increasing complexity.  
Each systems is solved via the pseudospectral discretisation as described in section §\ref{sec:solver}, producing a solution by time-integration 
using the Euler-forward method with timestep of $\Delta t = 5 \times 10^{-3}$. The time step is chosen to satisfy of the Courant-Friedrichs-Lewy
(CFL) condition to promote numerical stability. 
The model training is performed with  $1024$ training samples and $256$ validation samples, which are pseudo-randomly selected from the 
time-domain of the solution.  The \textit{adam} optimiser is employed~\cite{kingma2015} 
with a learning rate of $3 \times 10^{-4}$. The weighting factor for the loss $\mathcal{L}_{\bm \theta}$ (Eq.~\ref{eqn:optimization_problem}) is 
$\alpha = 10^{3}$, which is empirically determines to provide stable training. Each model is trained for a total of $10^4$ epochs, which is chosen to 
provide sufficient convergence. The accuracy of prediction is quantified by the relative error on the validation dataset 
\begin{equation} \label{eqn:error}
    e = 
    \sqrt{
        \frac{
            \sum_{t \in \mathcal{T}} \lVert {\bm u}^\ast(\bm{\Omega_g}, t) - {\bm {\eta_\theta}}({\bm \zeta}(\bm{\Omega_g}, t)) \rVert_{\bm{\Omega_g}}^{2}
        }{
            \sum_{t \in \mathcal{T}} \lVert {\bm u}^\ast(\bm{\Omega_g}, t) \rVert_{\bm{\Omega_g}}^{2}
        }
    }.
\end{equation}
This metric takes the magnitude of the solution into account, which allows the results from different systems to be compared.
All experiments are run on a single Quadro RTX 8000 GPU. 

\subsection{Uncovering solutions from corrupted data}
In this section, we uncover solutions from corrupted data with testcases on the linear convection-diffusion equation, the 
Burgers' equation, and the two-dimensional turbulent Navier-Stokes flow.

\subsubsection{The linear convection-diffusion equation} \label{sec:qualitative:linear}
The linear convection-diffusion equation is used to describe a variety of physical phenomena~\cite{majda1999simplified}. The equation is defined as 
\begin{equation} \label{eqn:linear_cd}
    \partial_t {\bm u} + {\bm c}\cdot \nabla {\bm u} = {\nu} \Delta {\bm u}, 
\end{equation}
where $\nabla $ is the nabla operator; $\Delta$ is the Laplacian operator; ${\bm c}\equiv (c,c)$, where $c$ is the convective
coefficient; and ${\nu}$ is the diffusion coefficient. The dissipative nature of the flow is further exacerbated by the presence
of periodic boundary conditions. The flow energy is subject to rapid decay because the solution quickly converge towards a fixed-point
solution at ${\bm u}^\ast(\bm{\Omega_g}, t) = 0$. In the presence of the fixed-point solution, we observe ${\bm \zeta}({\bm \Omega}_g, t) = {\bm \phi}({\bm \Omega_g})$,
which is a trivial case for identification and removal of the systematic error. In order to avoid rapid convergence to the fixed-point solution,
we take ${c} = 1.0, {\nu} = \nicefrac{1}{500}$ as the coefficients, producing a convective-dominant solution. 

A snapshot of the results for $k_\phi = 3, \mathcal{M} = 0.5$ is shown in Figure \ref{fig:linear_convection_diffusion}. There is a marked
difference between the corrupted observations ${\bm \zeta}({\bm \Omega}_{g}, t)$ in panel (i) and the underlying solution ${\bm u}^\ast({\bm \Omega}_{g}, t)$ 
in panel (ii), most notably in the magnitude of the field. Network predictions ${\bm {\eta_\theta}}({\bm \zeta}({\bm \Omega}_{g}, t))$ in 
panel (iii) uncover the underlying solution to the partial differential equation with a relative error of $e=6.612 \times 10^{-2}$ on the
validation set. 

\subsubsection{Nonlinear convection-diffusion equation} \label{sec:qualitative:nonlinear}
The nonlinear convection-diffusion equation, also known as Burgers' equation~\cite{burgers1948}, is  
\begin{equation} \label{eqn:nonlinear_cd}
    \partial_t {\bm u} + \left( {\bm u} \cdot {\nabla} \right) {\bm u} = {\nu} \Delta {\bm u},
\end{equation}
where the nonlinearity is in the convective term $\left( {\bm u} \cdot {\nabla} \right) {\bm u}$, the cause of turbulence in 
the Navier-Stokes equations. The nonlinear convective term provides a further challenge: below a certain threshold, the 
velocity interactions lead to further energy decay. The kinematic viscosity is set to ${\nu} = \nicefrac{1}{500}$, which produces
a convective-dominant solution by time integration.

In contrast to the linear convection-diffusion system, the relationship between the dynamics of corrupted-state and the true-state is 
more challenging. The introduction of nonlinearities in the differential operator revoke the linear relationship between the systematic error and
observed state, i.e., $\mathcal{R}({{\bm \zeta}({\bm x}, t)}; \lambda) \neq \mathcal{R}({{\bm \phi}({\bm x}, t)}; \lambda)$ 
as discussed in section §\ref{sec:defining_corruption_removal}. Consequently, this increases the complexity of the residual-based loss term $\mathcal{L}_{\mathcal{R}}$.
Figure \ref{fig:nonlinear_convection_diffusion} shows a snapshot of results for $k_\phi = 5, \mathcal{M} = 0.5$. The underlying solution
${\bm u}^\ast({\bm \Omega}_{g}, t)$ of the partial differential equation contains high-frequency spatial structures, shown in panel (ii), which
are less prominent in the corrupted data ${\bm \zeta}({\bm \Omega}_{g}, t)$, shown in panel (i). Despite the introduction of a nonlinear 
differential operator, we demonstrate that the network retains the ability to recover the underlying solution, with no notable difference 
in the structure between the predicted and ground-truth fields, shown in panels (iv) and (ii), respectively. The relative error on the 
validation set is $e=6.791 \times 10^{-2}$.

\subsubsection{Two-dimensional turbulent flow} \label{sec:qualitative:kolmogorov}
We consider a chaotic flow, which is governed by the incompressible Navier-Stokes equations evaluated on a two-dimensional domain with 
periodic boundary conditions and a periodic forcing term. This flow is also knows as the Kolmogorov flow~\cite{Fylladitakis2018}. The 
equations are expressions of the mass and momentum conservation, respectively 
\begin{equation} \label{eqn:kolmogorov}
\begin{aligned}
    {\nabla} \cdot {\bm u} &= 0, \\
    \partial_t {\bm u} + \left( {\bm u} \cdot {\nabla} \right) {\bm u} &= - {\nabla} p + { \nu} \Delta {\bm u} + {\bm g}.
\end{aligned}
\end{equation}
where $p$ is the scalar pressure field; ${\nu}$ is the kinematic viscosity; and ${\bm g}$ is a body forcing, which enables
the dynamics to be sustained by ensuring that the flow energy does not  dissipate at regime. The flow density is constant and assumed to be unity.

The use of a pseudospectral discretisation allows us to eliminate the pressure term and handle the continuity constraint implicitly, as shown in
the spectral representation of the two-dimensional turbulent flow
\begin{equation}
\begin{aligned}
    \left( \frac{d}{dt} + \nu \lvert \bm{k} \rvert^2 \right) \bm{\hat{u}}_k &= \bm{\hat{f}}_k - \bm{k} \frac{\bm{k} \cdot \bm{\hat{f}}_k}{\lvert \bm{k} \rvert^2} +\bm{\hat{g}}_k
    \qquad
        \text{with}
            \quad \bm{\hat{f}}_k = - \left( \widehat{\bm{u} \cdot \nabla \bm{u}} \right)_{k}, \\
    \mathcal{R}(\bm{\hat{u}}_k; \lambda) &= \left( \frac{d}{dt} + \nu \lvert \bm{k} \rvert^2 \right) \bm{\hat{u}}_k - \bm{\hat{f}}_k + \bm{k} \frac{\bm{k} \cdot \bm{\hat{f}}_k}{\lvert \bm{k} \rvert^2} - \bm{\hat{g}}_k,
\end{aligned}
\end{equation}
where nonlinear terms are handled in the pseudospectral sense, employing the standard $\nicefrac{2}{3}$ de-aliasing rule to avoid un-physical
culmination of energy at the high frequencies \cite{Canuto1988}. As a result of implicit handling of the continuity constraint, we can use the
residual-based loss as prescribed without imposing any further constraints for mass conservation.

The presence of the pressure gradient, ${\bm \nabla} p$, and forcing term, ${\bm g} = {\bm g}({\bm x})$ induce chaotic dynamics in the form of turbulence
for $Re = \nicefrac{1}{\nu} > 40$. We take ${\nu} = \nicefrac{1}{42}$ to ensure chaotic dynamics, imposing constant periodic forcing with 
${\bm g}(\bm{x}) = (\sin{(4\bm{x_2}), 0})$. In order to remove the transient and focus on a statistically stationary regime, the transient up to $T_t = 180.0$ 
is not recorded in the data.

\begin{figure}
    \centering
    \begin{subfigure}[h]{0.9\linewidth}
        \centering
        \includegraphics[width=\linewidth]{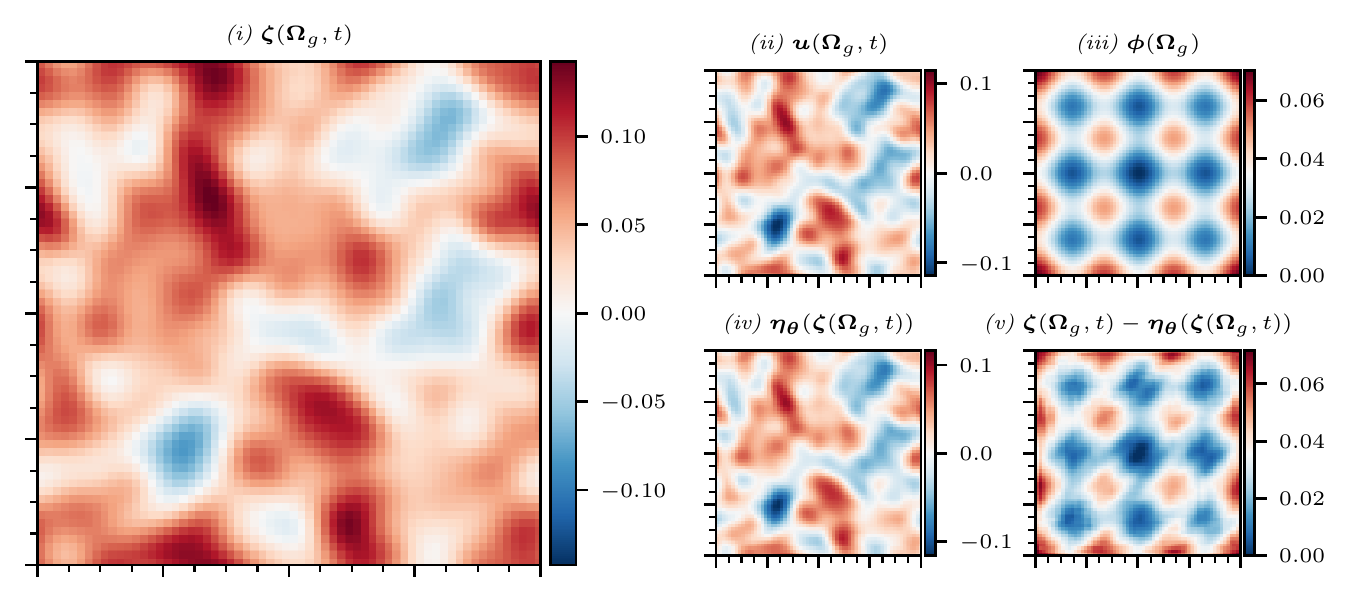}
        \caption{Linear convection-diffusion case $[k_\phi = 3, \mathcal{M}=0.5]$.}
        \label{fig:linear_convection_diffusion}
    \end{subfigure}
    \begin{subfigure}[h]{0.9\linewidth}
        \centering
        \includegraphics[width=\linewidth]{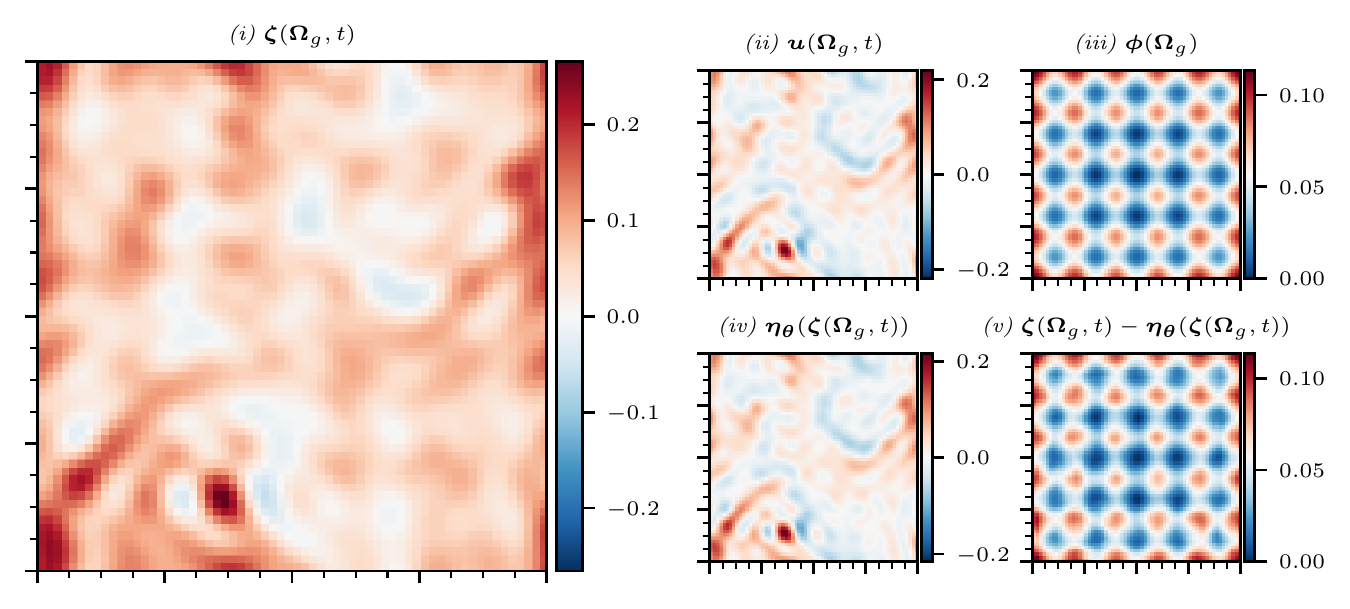}
        \caption{Nonlinear convection-diffusion case $[k_\phi = 5, \mathcal{M}=0.5]$.}
        \label{fig:nonlinear_convection_diffusion}
    \end{subfigure}
    \begin{subfigure}[h]{0.9\linewidth}
        \centering
        \includegraphics[width=\linewidth]{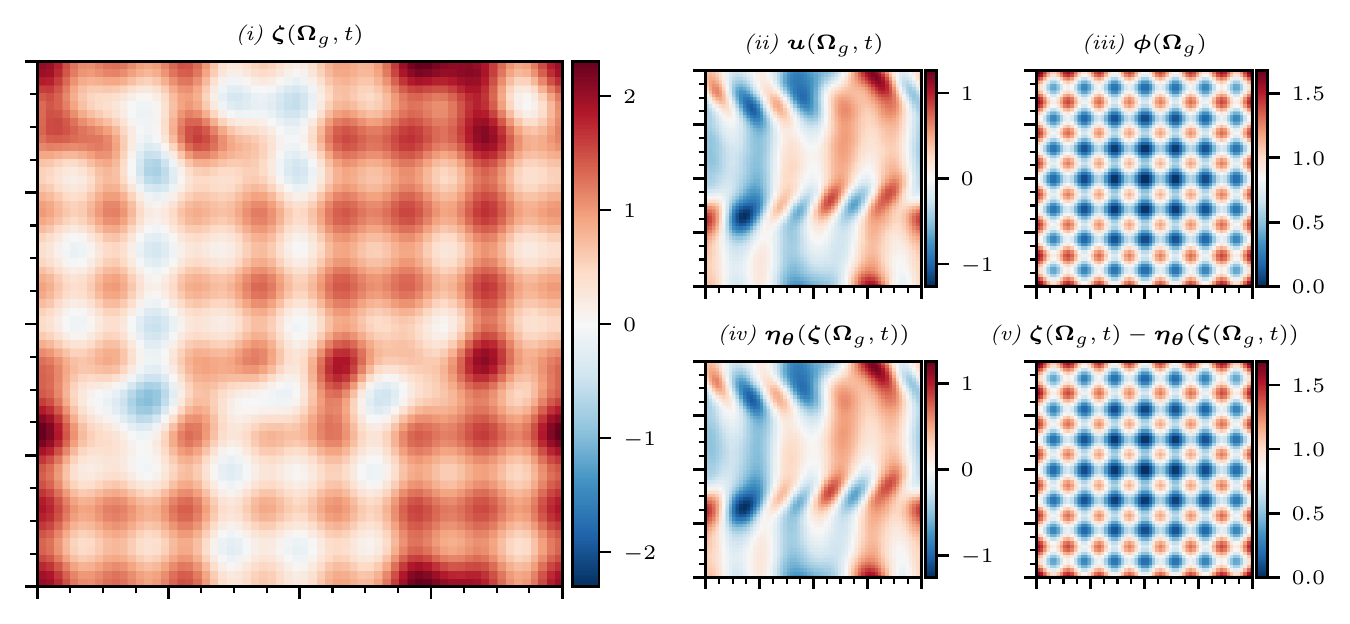}
        \caption{Two-dimensional turbulent flow case $[k_\phi = 7, \mathcal{M}=0.5]$.}
        \label{fig:kolmogorov}
    \end{subfigure}
        \caption{
            Snapshots of uncovered solutions from corrupted data. Panel (i) shows the corrupted field; (ii) shows the ground-truth solution; 
            (iii) shows the true systematic error; and (iv) shows the network predictions; (v) shows the predicted systematic error.
        }
        \label{fig:snapshot_results}
\end{figure}

A snapshot for the two-dimensional turbulent flow case is shown in Figure \ref{fig:kolmogorov} for $k_\phi = 7, \mathcal{M} = 0.5$. The corrupted
field ${\bm \zeta}({\bm \Omega}_{g}, t)$ as shown in panel (i) contains prominent, high-frequency spatial structures not present in the underlying 
solution ${\bm u}^\ast({\bm \Omega}_{g}, t)$ shown in panel (ii). The corrupted field bares little resemblance to the true solution. In spite of the
chaotic dynamics of the system, we demonstrate that the network  ${\bm {\eta_\theta}}({\bm \zeta}({\bm \Omega}_{g}, t))$ shown in panel
(iv) successfully removes the systematic error. The relative error on the  validation set is $e=2.044 \times 10^{-2}$.
%
\FloatBarrier
\subsection{Robustness} \label{sec:numerical_studies}
In this section, we analyse the robustness of the methodology to systematic errors of varying modality and magnitude.
The systematic error, ${\bm \phi}(\bm{\Omega}_g)$, is varied for a range of Rastrigin wavenumbers $k_\phi$ and relative magnitudes $\mathcal{M}$. 
These parameter ranges allow us to assess the degree to which the underlying solution of a partial differential equation can be recovered 
when subjected to spatially varying systematic error with different degrees of multi-modality and magnitudes. To this end, we propose 
two parametric studies for each partial differential equation
\begin{equation*}
\begin{aligned}[c]
    &\mathit{(i)} \\
    &\mathit{(ii)}
\end{aligned}
\quad
\begin{aligned}[c]
    &\mathcal{M} = 0.5, \\
    &k_\phi = 3,
\end{aligned}
\quad
\begin{aligned}[c]
    k_\phi &\in \{ 1, 3, 5, 7, 9 \}; \\
    \mathcal{M} &\in \{ 0.01, 0.1, 0.25, 0.5, 1.0 \}.
\end{aligned}
\end{equation*}
In the first case, we fix the relative magnitude and vary the Rastrigin wavenumber, whereas in the second case we fix the Rastrigin wavenumber
and vary the relative magnitude. For each case, we compute the relative error, $e$, between the predicted solution 
${\bm {\eta_\theta}}({\bm \zeta}(\bm{\Omega_g}, t))$ and the true solution ${\bm u}^\ast(\bm{\Omega}_g, t)$, as defined in Eq.~\eqref{eqn:error}. 
We show the mean error for five repeats of each experiment to ensure results are representative of the true performance. Empirically, we find that
performance is robust to pseudo-random initialisation of network parameters with standard deviation of $\mathcal{O}(10^{-4})$.

Computational resource is fixed for each run, which employs the same experimental setup described in section §\ref{sec:overview_results}. 
Whilst the optimal hyperparameters are likely to vary on a case-by-case basis, providing a common baseline allows for 
explicit comparison between cases. Assessing results using the same hyperparameters for the three partial differential equations allows us to explore
the robustness of the methodology.

\begin{figure}
    \centering
    \begin{subfigure}[b]{\linewidth}
        \centering
        \includegraphics[width=\linewidth]{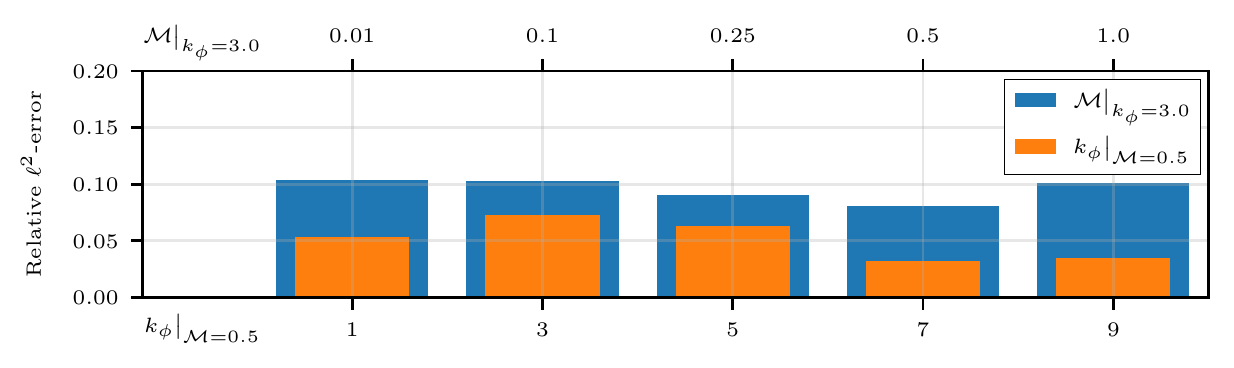}
        \caption{Linear convection-diffusion case.}
        \label{fig:p_linear_convection_diffusion}
    \end{subfigure}
    \hfill
    \begin{subfigure}[b]{\linewidth}
        \centering
        \includegraphics[width=\linewidth]{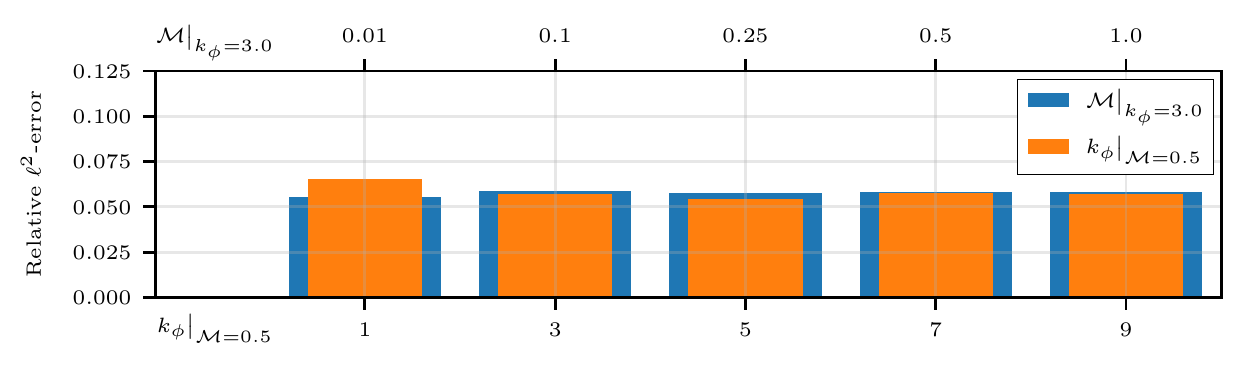}
        \caption{Nonlinear convection-diffusion case.}
        \label{fig:p_nonlinear_convection_diffusion}
    \end{subfigure}
    \begin{subfigure}[b]{\linewidth}
        \centering
        \includegraphics[width=\linewidth]{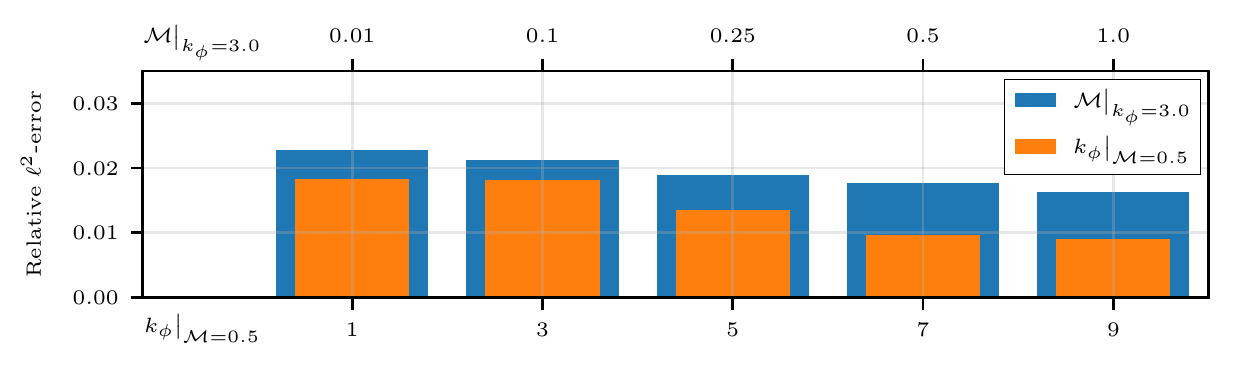}
        \caption{Two-dimensional turbulent flow case.}
        \label{fig:p_kolmogorov}
    \end{subfigure}
        \caption{
           Robustness analysis. Results for the parameterised studies for each of the three partial differential equations. 
            Orange-bars denote results for case $\mathit{(i)}$: fixing the magnitude and varying the Rastrigin wavenumber.
            Blue-bars denote results for case $\mathit{(ii)}$: fixing the Rastrigin wavenumber and varying the magnitude.
        }
        \label{fig:p_results}
\end{figure}

In the case of systematic error removal for the linear convection-diffusion problem, shown in Figure \ref{fig:p_linear_convection_diffusion},
we demonstrate that the relative error is largely independent of the form of parameterised systematic error. The relative magnitude
$\mathcal{M}$ and Rastrigin wavenumber $k_\phi$ have little impact on the ability of the model to uncover the true solution to the partial differential
equation, with performance consistent across all cases. The model performs best for the case $k_\phi = 7, \mathcal{M} = 0.5$, achieving 
a relative error of $2.568 \times 10^{-1}$.  

Results for the nonlinear convection-diffusion case show marked improvement in comparison with the linear case, with errors for the
numerical study shown in Figure \ref{fig:p_nonlinear_convection_diffusion}. Despite the introduction of a nonlinear operator, the 
relative error is consistently lower regardless of the parameterisation of systematic error. The network remains equally capable 
of uncovering the underlying solution across a wide range of modalities and magnitudes. 

Although the residual-error from the network predictions ${\bm {\eta_\theta}}({\bm \zeta}(\bm{\Omega}_g, t))$ no longer scales in a linear fashion, this
is beneficial for the training dynamics. This is because the second order nonlinearity promotes convexity in the loss-surface, which is then
exploited by our gradient-based optimisation approach. 

Introduction of chaotic dynamics in the case of the two-dimensional turbulent flow does not have adverse effects on the ability of the proposed method  
to uncover the underlying solution of the partial differential equation. Results shown in Figure \ref{fig:p_kolmogorov} demonstrate an
improvement in performance from the standard nonlinear convection-diffusion case. For both fixed Rastrigin wavenumber 
and magnitude, increasing the value of the parameter tends to decrease  the relative error. In other words, the method is not influenced by the complexity of the partial differential equation used in the loss calculations. 
%

\FloatBarrier

\section{Physical consistency of uncovered Navier-Stokes solutions}

Results in section §\ref{sec:numerical_studies}  show  that, for a generic training setup, we are able to achieve low values of the relative 
error regardless of modality or magnitude of systematic error. Because the two-dimensional turbulent flow case is chaotic, infinitesimal 
errors $\mathcal{O}({\epsilon})$ exponentially grow in time, therefore, the residual $\mathcal{R}({\bm u}^\ast(\bm{\Omega}_g, t) + { \epsilon}(\bm{\Omega}_g, t); \lambda) \gg \mathcal{R}({\bm u}^\ast(\bm{\Omega}_g, t); \lambda)$ 
where ${ \epsilon}$ is a perturbation parameter. In this section, we analyse the physical properties of the solutions of the Navier-Stokes equation
(Eq.~\ref{eqn:kolmogorov}) uncovered by the PC-CNN, ${\bm {\eta_\theta}}({\bm \zeta}(\bm{\Omega}_g, t)))$. 

\begin{figure}[ht]
    \centering
    \includegraphics[width=\linewidth]{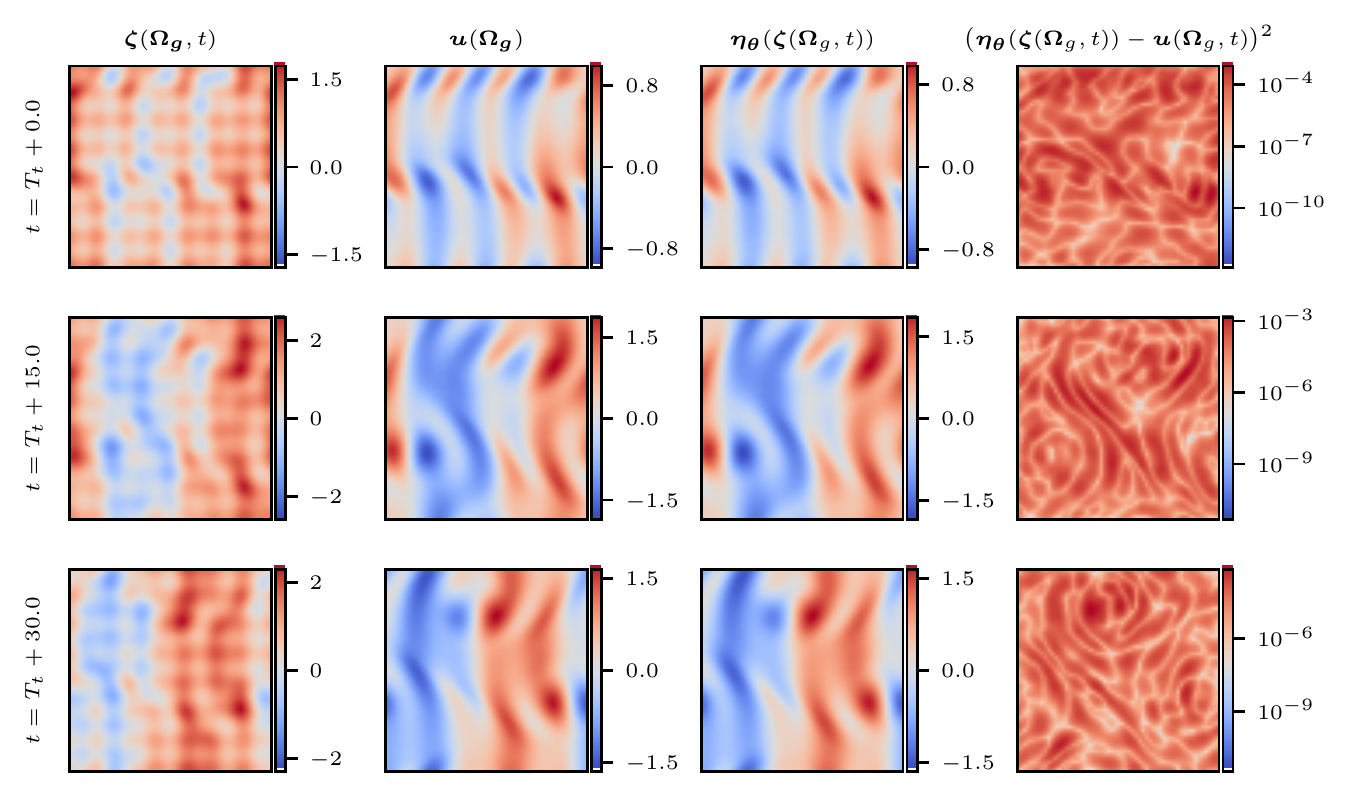}
    \caption{
        Systematic error removal for the two-dimensional turbulent flow $[k_\phi = 7, \mathcal{M} = 0.5]$. $T_t$ denotes the length of the transient.
    }
    \label{fig:ts_kolmogorov}
\end{figure}

\begin{figure}[ht]
    \centering
    \includegraphics[width=\linewidth]{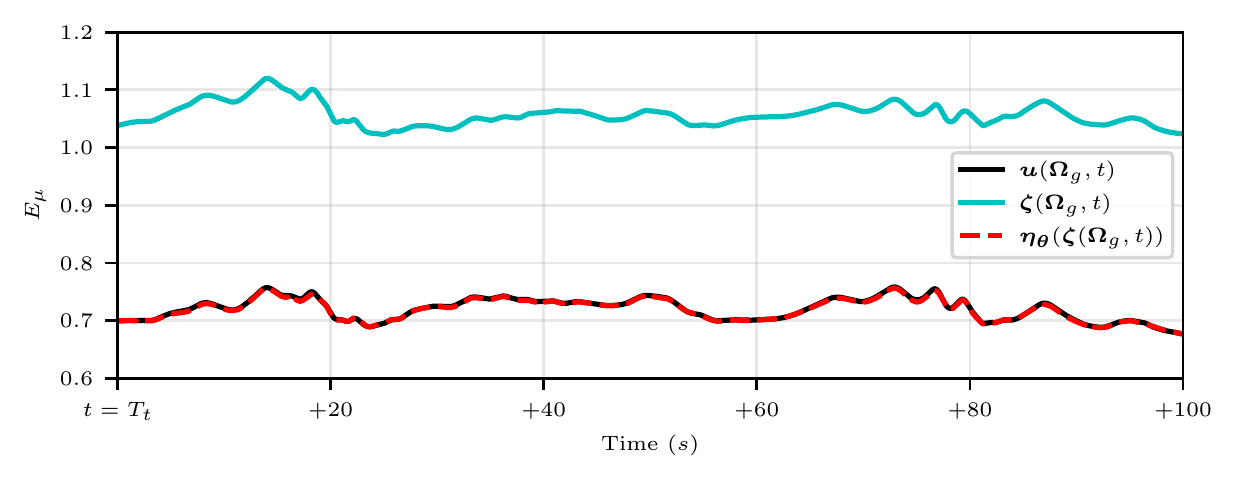}
    \caption{Kinetic energy for the two-dimensional turbulent flow $[k_\phi=7, \mathcal{M}=0.5]$.}
    \label{fig:ke_plot}
\end{figure}

\begin{figure}[ht]
    \centering
    \includegraphics[width=\linewidth]{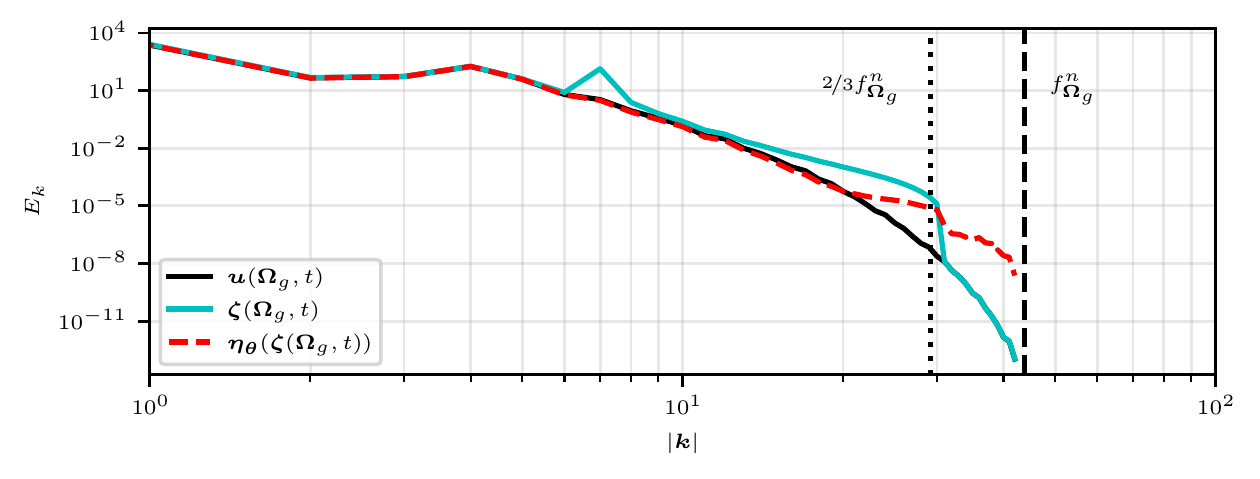}
    \caption{
        Energy Spectrum for the two-dimensional turbulent flow $[k_\phi = 7, \mathcal{M} = 0.5]$.
    }
    \label{fig:es_kolmogorov}
\end{figure}

First, we show snapshots of the time evolution of the two-dimensional turbulent flow in Figure \ref{fig:ts_kolmogorov}. These confirms that the model is 
learning a physical solution across the time-domain of the solution, as shown in the error on a log-scale in the final column. The parameterisation
of the systematic error is fixed in this case with $k_\phi = 7, \mathcal{M} = 0.5$.
Second, we analyse the statistics of the solution. The mean kinetic energy of the flow at each timestep is directly affected by the 
introduction of the systematic error. In the case of our strictly positive systematic error ${\bm \phi}(\bm{\Omega}_g)$, the mean kinetic energy 
is increased at every point. Results in Figure \ref{fig:ke_plot} show the kinetic energy time series across the time-domain for (i) the
ground truth ${\bm u}^\ast(\bm{\Omega}_g, t)$; the corrupted data ${\bm \zeta}(\bm{\Omega}_g, t)$; and the network's predictions 
${\bm {\eta_\theta}}({\bm \zeta}(\bm{\Omega}_g, t))$. The chaotic nature of the solution can be observed from the spikes in chaotic energy. 
We observe that the network predictions accurately reconstruct the kinetic energy trace across the simulation domain, despite fewer than
$2\%$ of the available timesteps being used for training.
Third, we analyse the energy spectrum, which is characteristic of turbulent flows, in which the energy content decreases with the wavenumber. Introducing the systematic error at a 
particular wavenumber artificially alters the energy spectrum due to increased energy content. In Figure \ref{fig:es_kolmogorov}, we 
show the energy spectrum for the two-dimensional turbulent flow, where this unphysical spike in energy content is visible for $\lvert \bm{k} \rvert \geq 7$.
Model predictions ${\bm {\eta_\theta}}({\bm \zeta}(\bm{\Omega}_g, t))$ correct the energy content for $\lvert \bm{k} \rvert < 21$, and successfully 
characterise and reproduce scales of turbulence. The exponential decay of energy content with increasing spatial frequency provides
a significant challenge for the residual-based loss. Despite the challenge, we observe only minor discrepancy from the true solution at large wavenumbers, i.e., when aliasing occurs ($\lvert \bm{k} \rvert \gtrapprox 29$).
The ability of the network ${\bm {\eta_\theta}}$ to recover the underlying solution from corrupted data is demonstrated to yield low-error 
when compared with the true underlying solution of the partial differential equation. By investigating properties and statistics of the
solution we demonstrate that, even for turbulent nonlinear systems, we are able to produce predictions that adhere to the physical properties
of the system.

\FloatBarrier

\section{Conclusion} \label{sec:conclusions}

Data and models can be corrupted by systematic errors, which may originate from miscalibration of the experimental apparatus and epistemic uncertainties.  
We introduce a methodology to remove the systematic error from data to uncover the physical solution of the partial differential equation.
First, we introduce the physics-constrained convolutional neural network (PC-CNN), which provides the means to compute the physical 
residual, i.e., the areas of the field in which the physical laws are violated because of the systematic error. 
Second, we formulate an optimisation problem by leveraging prior knowledge of the underlying
physical system to regularise the predictions from the PC-CNN. 
Third, we numerically test the method to remove systematic errors from data produced by
three partial differential equations of increasing complexity (linear and nonlinear diffusion-convection, and Navier-Stokes). This shows the ability of the PC-CNN to accurately recover the underlying solutions. 
Fourth, we carry out a parameterised analysis, which successfully minimises the relative error for 
a variety of systematic errors with different magnitudes and degrees of multimodality. This shows that the method is robust. Finally, we investigate the physical consistency of the inferred solutions for the two-dimensional turbulent flow.  The network predictions satisfy physical properties of the underlying partial differential equation (Navier-Stokes), such as the  statistics and energy spectrum. 
This work opens opportunities for inferring solutions of partial differential equations from sparse measurements. 
Current work is focused on  scaling up the framework to three-dimensional flows, and dealing with experimental data.

\section*{Acknowledgements}
D. Kelshaw. and L. Magri. acknowledge support from the UK Engineering and Physical Sciences Research Council. 
L. Magri gratefully acknowledges financial support from the ERC Starting Grant PhyCo 949388.

\bibliography{references}

\begin{thebibliography}{37}
\providecommand{\natexlab}[1]{#1}
\providecommand{\url}[1]{#1}
\csname url@samestyle\endcsname
\providecommand{\newblock}{\relax}
\providecommand{\bibinfo}[2]{#2}
\providecommand{\BIBentrySTDinterwordspacing}{\spaceskip=0pt\relax}
\providecommand{\BIBentryALTinterwordstretchfactor}{4}
\providecommand{\BIBentryALTinterwordspacing}{\spaceskip=\fontdimen2\font plus
\BIBentryALTinterwordstretchfactor\fontdimen3\font minus
  \fontdimen4\font\relax}
\providecommand{\BIBforeignlanguage}[2]{{%
\expandafter\ifx\csname l@#1\endcsname\relax
\typeout{** WARNING: IEEEtranN.bst: No hyphenation pattern has been}%
\typeout{** loaded for the language `#1'. Using the pattern for}%
\typeout{** the default language instead.}%
\else
\language=\csname l@#1\endcsname
\fi
#2}}
\providecommand{\BIBdecl}{\relax}
\BIBdecl

\bibitem[Sciacchitano et~al.(2015)Sciacchitano, Neal, Smith, Warner, Vlachos,
  Wieneke, and Scarano]{sciacchitano2015collaborative}
A.~Sciacchitano, D.~R. Neal, B.~L. Smith, S.~O. Warner, P.~P. Vlachos,
  B.~Wieneke, and F.~Scarano, ``Collaborative framework for piv uncertainty
  quantification: comparative assessment of methods,'' \emph{Measurement
  Science and Technology}, vol.~26, no.~7, p. 074004, 2015.

\bibitem[Zucchelli et~al.(2021)Zucchelli, Delande, Jones, and
  Jah]{zucchelli2021multi}
E.~M. Zucchelli, E.~D. Delande, B.~A. Jones, and M.~K. Jah, ``Multi-fidelity
  orbit determination with systematic errors,'' \emph{The Journal of the
  Astronautical Sciences}, vol.~68, pp. 695--727, 2021.

\bibitem[Adrian(1991)]{adrian1991particle}
R.~J. Adrian, ``Particle-imaging techniques for experimental fluid mechanics,''
  \emph{Annual review of fluid mechanics}, vol.~23, no.~1, pp. 261--304, 1991.

\bibitem[Vedula and Adrian(2005)]{Vedula2005}
P.~Vedula and R.~J. Adrian, ``Optimal solenoidal interpolation of turbulent
  vector fields: Application to ptv and super-resolution piv,''
  \emph{Experiments in Fluids}, vol.~39, pp. 213--221, 8 2005.

\bibitem[Schiavazzi et~al.(2014)Schiavazzi, Coletti, Iaccarino, and
  Eaton]{Schiavazzi2014}
D.~Schiavazzi, F.~Coletti, G.~Iaccarino, and J.~K. Eaton, ``A matching pursuit
  approach to solenoidal filtering of three-dimensional velocity
  measurements,'' \emph{Journal of Computational Physics}, vol. 263, pp.
  206--221, 4 2014.

\bibitem[Kochkov et~al.(2021)Kochkov, Smith, Alieva, Wang, Brenner, and
  Hoyer]{kochkov2021machine}
D.~Kochkov, J.~A. Smith, A.~Alieva, Q.~Wang, M.~P. Brenner, and S.~Hoyer,
  ``Machine learning--accelerated computational fluid dynamics,''
  \emph{Proceedings of the National Academy of Sciences}, vol. 118, no.~21, p.
  e2101784118, 2021.

\bibitem[Kumar and Nachamai(2017)]{Kumar2017}
N.~Kumar and M.~Nachamai, ``Noise removal and filtering techniques used in
  medical images,'' \emph{Oriental journal of computer science and technology},
  vol.~10, pp. 103--113, 3 2017.

\bibitem[Raiola et~al.(2015)Raiola, Discetti, and Ianiro]{Raiola2015}
M.~Raiola, S.~Discetti, and A.~Ianiro, ``On piv random error minimization with
  optimal pod-based low-order reconstruction,'' \emph{Experiments in Fluids},
  vol.~56, 4 2015.

\bibitem[Mendez et~al.(2017)Mendez, Raiola, Masullo, Discetti, Ianiro,
  Theunissen, and Buchlin]{Mendez2017}
M.~A. Mendez, M.~Raiola, A.~Masullo, S.~Discetti, A.~Ianiro, R.~Theunissen, and
  J.~M. Buchlin, ``Pod-based background removal for particle image
  velocimetry,'' \emph{Experimental Thermal and Fluid Science}, vol.~80, pp.
  181--192, 1 2017.

\bibitem[Vincent et~al.(2008)Vincent, Larochelle, Bengio, and
  Manzagol]{Vincent2008}
P.~Vincent, H.~Larochelle, Y.~Bengio, and P.-A. Manzagol, ``Extracting and
  composing robust features with denoising autoencoders,'' \emph{Proceedings of
  the 25th International Conference on Machine Learning}, pp. 1096--1103, 2008.

\bibitem[Freeman et~al.(2002)Freeman, Jones, and Pasztor]{Freeman2002}
W.~T. Freeman, T.~R. Jones, and E.~C. Pasztor, ``Example-based
  super-resolution,'' \emph{IEEE Computer Graphics and Applications}, vol.~22,
  pp. 56--65, 2002.

\bibitem[Yang et~al.(2010)Yang, Wright, Huang, and Ma]{Yang2010}
J.~Yang, J.~Wright, T.~S. Huang, and Y.~Ma, ``Image super-resolution via sparse
  representation,'' \emph{IEEE Transactions on Image Processing}, vol.~19, pp.
  2861--2873, 11 2010.

\bibitem[Dong et~al.(2014)Dong, Loy, He, and Tang]{Dong2014}
\BIBentryALTinterwordspacing
C.~Dong, C.~C. Loy, K.~He, and X.~Tang, ``Learning a deep convolutional network
  for image super-resolution,'' \emph{Computer Vision -- ECCV 2014184}, pp.
  184--199, 2014. [Online]. Available:
  \url{http://mmlab.ie.cuhk.edu.hk/projects/SRCNN.html.}
\BIBentrySTDinterwordspacing

\bibitem[Xie et~al.(2018)Xie, Franz, Chu, and Thuerey]{Xie2018}
Y.~Xie, E.~Franz, M.~Chu, and N.~Thuerey, ``Tempogan: A temporally coherent,
  volumetric gan for super-resolution fluid flow,'' \emph{ACM Transactions on
  Graphics}, vol.~37, 2018.

\bibitem[Hornik et~al.(1989)Hornik, Stinchcombe, and White]{Hornik1989}
K.~Hornik, M.~Stinchcombe, and H.~White, ``Multilayer feedforward networks are
  universal approximators,'' \emph{Neural Networks}, vol.~2, pp. 359--366,
  1989.

\bibitem[Zhou(2020)]{Zhou2020}
D.~X. Zhou, ``Universality of deep convolutional neural networks,''
  \emph{Applied and Computational Harmonic Analysis}, vol.~48, pp. 787--794, 3
  2020.

\bibitem[Lagaris et~al.(1998)Lagaris, Likas, and Fotiadis]{Lagaris1998}
I.~E. Lagaris, A.~Likas, and D.~I. Fotiadis, ``Artificial neural networks for
  solving ordinary and partial differential equations,'' \emph{IEEE
  Transactions on Neural Networks}, vol.~9, pp. 987--1000, 1998.

\bibitem[Raissi et~al.(2019)Raissi, Perdikaris, and Karniadakis]{Raissi2019}
M.~Raissi, P.~Perdikaris, and G.~E. Karniadakis, ``Physics-informed neural
  networks: A deep learning framework for solving forward and inverse problems
  involving nonlinear partial differential equations,'' \emph{Journal of
  Computational Physics}, vol. 378, pp. 686--707, 2 2019.

\bibitem[Rastrigin(1974)]{Rastrigin1974}
L.~A. Rastrigin, ``Systems of extremal control,'' \emph{Nauka}, 1974.

\bibitem[N{\'o}voa and Magri(2022)]{novoa2022real}
A.~N{\'o}voa and L.~Magri, ``Real-time thermoacoustic data assimilation,''
  \emph{Journal of Fluid Mechanics}, vol. 948, p. A35, 2022.

\bibitem[LeCun et~al.(2015)LeCun, Bengio, and Hinton]{lecun2015deep}
Y.~LeCun, Y.~Bengio, and G.~Hinton, ``Deep learning,'' \emph{nature}, vol. 521,
  no. 7553, pp. 436--444, 2015.

\bibitem[Gu et~al.(2018)Gu, Wang, Kuen, Ma, Shahroudy, Shuai, Liu, Wang, Wang,
  Cai, and Chen]{Gu2018}
\BIBentryALTinterwordspacing
J.~Gu, Z.~Wang, J.~Kuen, L.~Ma, A.~Shahroudy, B.~Shuai, T.~Liu, X.~Wang,
  G.~Wang, J.~Cai, and T.~Chen, ``Recent advances in convolutional neural
  networks,'' \emph{Pattern Recognition}, vol.~77, pp. 354--377, 2018.
  [Online]. Available:
  \url{https://www.sciencedirect.com/science/article/pii/S0031320317304120}
\BIBentrySTDinterwordspacing

\bibitem[Pope and Pope(2000)]{pope2000turbulent}
S.~B. Pope and Pope, \emph{Turbulent flows}.\hskip 1em plus 0.5em minus
  0.4em\relax Cambridge university press, 2000.

\bibitem[Racca et~al.(2022)Racca, Doan, and Magri]{racca2022modelling}
A.~Racca, N.~A.~K. Doan, and L.~Magri, ``Modelling spatiotemporal turbulent
  dynamics with the convolutional autoencoder echo state network,'' 2022.

\bibitem[Doan et~al.(2021)Doan, Polifke, and Magri]{doan2021}
N.~A.~K. Doan, W.~Polifke, and L.~Magri, ``Short and long-term predictions of
  chaotic flows and extreme events: a physics-constrained reservoir computing
  approach,'' \emph{Proceedings of the Royal Society A: Mathematical, Physical
  and Engineering Sciences}, vol. 477, no. 2253, p. 20210135, 2021.

\bibitem[Magri(2023)]{magri2023notes}
L.~Magri, ``{Introduction to neural networks for engineering and computational
  science},'' 1 2023.

\bibitem[Murata et~al.(2020)Murata, Fukami, and Fukagata]{Murata2020}
T.~Murata, K.~Fukami, and K.~Fukagata, ``Nonlinear mode decomposition with
  convolutional neural networks for fluid dynamics,'' \emph{Journal of Fluid
  Mechanics}, vol. 882, p. A13, 2020.

\bibitem[Gao et~al.(2021)Gao, Sun, and Wang]{Gao2021}
H.~Gao, L.~Sun, and J.-X. Wang, ``Super-resolution and denoising of fluid flow
  using physics-informed convolutional neural networks without high-resolution
  labels,'' \emph{Physics of Fluids}, vol.~33, p. 073603, 7 2021.

\bibitem[Bishop(1994)]{Bishop1994}
C.~M. Bishop, ``Neural networks and their applications,'' \emph{Review of
  Scientific Instruments}, vol.~65, no.~6, pp. 1803--1832, 1994.

\bibitem[Aloysius and Geetha(2017)]{Aloysius2017}
N.~Aloysius and M.~Geetha, ``A review on deep convolutional neural networks,''
  pp. 0588--0592, 2017.

\bibitem[Kelshaw(2022)]{kolsol2022}
D.~Kelshaw, ``Kolsol,'' \url{https://github.com/magrilab/kolsol}, 2022.

\bibitem[Canuto et~al.(1988)Canuto, Hussaini, Quarteroni, and Zang]{Canuto1988}
C.~Canuto, M.~Y. Hussaini, A.~Quarteroni, and T.~A. Zang, \emph{Spectral
  Methods in Fluid Dynamics}.\hskip 1em plus 0.5em minus 0.4em\relax Springer
  Berlin Heidelberg, 1988.

\bibitem[Bateman(1915)]{bateman1915}
H.~Bateman, ``Some recent researches on the motion of fluids,'' \emph{Monthly
  Weather Review}, vol.~43, no.~4, pp. 163 -- 170, 1915.

\bibitem[Burgers(1948)]{burgers1948}
J.~Burgers, ``A mathematical model illustrating the theory of turbulence,''
  vol.~1, pp. 171--199, 1948.

\bibitem[Fylladitakis(2018)]{Fylladitakis2018}
E.~D. Fylladitakis, ``Kolmogorov flow: Seven decades of history,''
  \emph{Journal of Applied Mathematics and Physics}, vol.~6, pp. 2227--2263,
  2018.

\bibitem[Kingma and Ba(2015)]{kingma2015}
D.~P. Kingma and J.~Ba, ``Adam: {A} method for stochastic optimization,'' 2015.

\bibitem[Majda and Kramer(1999)]{majda1999simplified}
A.~J. Majda and P.~R. Kramer, ``Simplified models for turbulent diffusion:
  theory, numerical modelling, and physical phenomena,'' \emph{Physics
  reports}, vol. 314, no. 4-5, pp. 237--574, 1999.

\end{thebibliography}

\end{document}